# Optimizing SAR data processing and thresholding for forest change detection: an application for early deforestation warnings on eastern Amazonia

*Juan Doblas Prieto[1]*

[1]Instituto Nacional de Pesquisas Espaciais (INPE), juan.doblas@inpe.br

**ABSTRACT**

*The present work proposes a prototype for an operational method for early deforestation detection of cloudy tropical rainforests. The proposed methodology makes use of Sentinel-1 SAR data processed into the Google Earth Engine platform for flag the areas where the probability of recent deforestation is high. The evaluation of the results over a region on the Eastern Amazon basin showed that copolarized data (VV band) offers the best results in terms of producer's accuracy (95,4% for a 5% significance, 88,9% for 1% significance), while crosspolarized data (VH band) offered excellent results in terms of user's accuracy (86% for a 5% significance, 100% for 1% significance).*

**RESUMO**

*O presente trabalho propõe um protótipo de uma metodologia operacional que visa a detecção de desmatamento em regiões de floresta tropical densa com cobertura frequente de nuvens. A metodologia usa dados Sentinel-1, processados na plataforma Google Earth Engine, para sinalizar os pixels que apresentam maior probabilidade de corresponder a áreas recentemente desmatadas. A avaliação dos resultados da detecção sobre uma região na Amazônia oriental mostrou que os dados copolarizados (VV) apresentaram a melhor acurácia do produtor (95,4% para um nível de significância de 5%, 88,9% para um nível de significância de 1%) e os dados de polarização cruzada (VH) ofereceram melhores resultados em termos da acurácia do usuário (86% para um nível de significância de 5%, 100% para um nível de significância de 1%).*

**Palavras-chave** — *SAR, Monitoramento de florestas, Desmatamento, alertas rápidas*

## 1. INTRODUCTION

Change detection (CD) is the process of identifying differences in the state of an object or phenomenon by observing it at different times [1]. In remote sensing, change detection tries to compare spatial representation of two or more points in time, measuring what changes are due to variable of interest (forest cover, water level, elevation) and controlling all variances caused by differences in variables that are not of interest [2].

There are a variety of factors that can affect the quality of a change detection research, being the two most important: 1) precise geometric registration and 2) precise calibration between multi-temporal images [3]. It can be said that for most of applications, precise co-registration is not a problem nowadays (being multi-source data fusion a notorious exception). Meanwhile, calibration can be a serious issue, especially on optical imagery. Passive sensors, such as those used by the Landsat satellite series, use a variable, incoherent source of energy (solar emission), which is, additionally, severely affected by atmospheric absorption and scattering. Active microwave systems use their own emitted radiation, which is affected by atmospheric conditions to a much lesser extent. That makes radar imaging sensors good candidates for change detection research on cloudy environments [4].

Orbital Synthetic Aperture Radar (SAR) data has been used to monitor changes on the earth surface since, at least, 1970 [5]. Nowadays, SAR change detection (SAR-CD) is used on many applications such as environmental monitoring, study on land-use/land-cover dynamics, analysis of forest or vegetation changes, damage assessment, agricultural surveys, and analysis of urban changes [6].

Radar data offers some advantages on change detection, and a few cons, if compared to the more popular, widely used optical data. The most obvious advantage is the ability of the SAR sensors to record surface information (almost) independently of weather and illumination conditions, as stated before. As this, effective monitoring of cloud-covered tropical forests or scarce illuminated polar regions can take advantage of SAR characteristics [7].

Speckle contamination of the SAR data is an intrinsic, central issue on SAR applications. The multiplicative nature of speckle makes it difficult to remove, as the noise statistical characteristics are quite close to the real-data ones, both in space and frequency domains. As this, every change detection technique using SAR data should address this problem before any data manipulation, as it can be extremely harmful to the accuracy of the results [8].

Also, and due to the intrinsic characteristics of the SAR method, deforestation detection on SAR data has proven to be a complex issue, as the backscattering signal change over deforestation patches can be quite different depending on the deforestation stage and type [9]. Some authors tried to avoid this problem focusing on the geometrical effects produced by deforestation, such as shadowing [10].

Regarding the recent use of SAR-CD techniques to produce early warnings of deforestation on tropical forests, the work of the university of Wageningen's Laboratory of Geo-information Science and Remote Sensing is remarkable, as it aims the construction of an operational, feasible workflow able to detect in the shortest span of time [11].

The objective of this work was to adapt the methods suggested by [11] to a region on western Amazonia, using existing, reliable, reference data, multi-band thresholding and an enhanced, cloud-based SAR signal processing workflow to achieve an optimal automatic deforestation detection rate on dense canopy areas.

## 2. METHODOLOGY

### 2.1. Area of Interest

The Area of Interest (AOI) covers a 67.000 km² tilted rectangle over the Brazilian eastern Amazon. The area follows a ~500 km stretch of the Transamazonian highway, between the cities of Altamira and Itaituba, on the Brazilian state of Pará (figure 1).

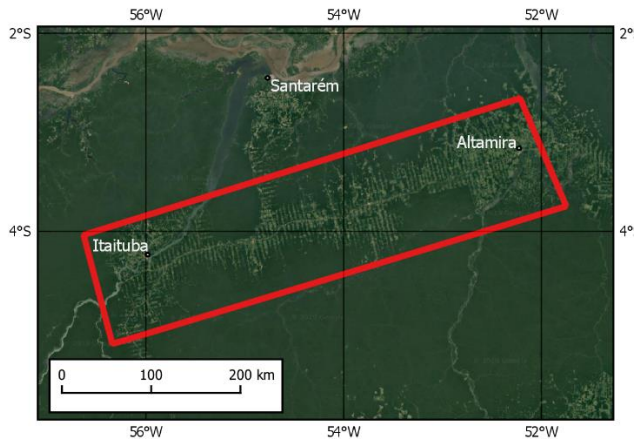

**Figure 1. Study area, located along the Transamazonian Highway (BR-230), between the cities of Altamira and Itaituba, on the Brazilian state of Pará. Background image: Google Inc.**

The AOI belongs to a tropical region, characterized a wet season from November to May, followed by a drier (but not completely dry) season, from June to October. Total precipitation in Altamira varies around 1800 mm/yr. Regarding vegetation, dense ombrophylous forest covers almost all the intact vegetation areas. Flooded forests are not present in the studied area.

The main drive of deforestation is land clearing for cattle-ranching in small to medium patches, corresponding to a low productivity, unsustainable production framed by land ownership concentration by largeholders [12]. Figure 2 shows a classic deforestation setup, with patches of forest being cleared on rectangular shapes next to previous deforested areas, along a secondary road.

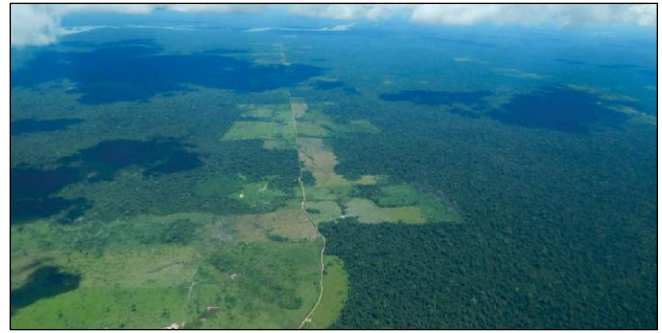

**Figure 2: Aerial image showing the spatial configuration of the deforestation patches along the secondary roads of the AOI. Source of the image: personal arquive.**

### 2.2. Reference data

Two different sampling sets were built for the purposes of the study: one *invariant forest data* sample set, made of 8 patches of intact forest area, as assured by the PRODES/INPE records (for review and download of PRODES and DETER data, see http://terrabrasilis.dpi.inpe.br/*)*. The *cleared forest data* sample test was built following a complex selection procedure:

1- The 2826 available polygons of the DETER/INPE dataset on the AOI were filtered to retain only clear cutted areas with no previous degradation.
2- The resulting 1203 polygon dataset was thinned, by removing all the areas which weren't observed, due to cloud cover, on the observation that predated the polygon creation date. This procedure, with retained 407 polygons, assured that the date of the polygon should be close to the date of the real deforestation.
3- Finally, the resulting polygons were intersected with the PRODES 2018 dataset. This final operation resulted on 196 fine-tuned, manually confirmed deforestation polygons.

### 2.3. Input SAR data

The data used on this study was Sentinel-1A SAR data. The Google Earth Engine platform (GEE, [13]) able one to visualize and process Sentinel-1A data on a distributed computing environment. The Sentinel-1 data made available on the GEE platform is already preprocessed, following the standard SNAP processing flow to obtain $\sigma^0$ backscattering values.

Sentinel-1 $\sigma^0$ data was converted to $\gamma^0$ using the local incidence angle (LIA), which was computed taking in account the local relief and the acquisition geometry at every pixel of the image.

### 2.4 Finding the optimal filter

Quegan & Yu temporal filtering (QY,[14]) has been applied with success over Sentinel-1A series ([10],[11]), and it can be considered nowadays as a fundamental part of the processing chain of dense temporal SAR data. The developers of this filter suggested that it can be advisable to [14]:
1. use an additional spatial filter after applying the temporal filter, and
2. use adaptative filters to compute the estimated speckle-free backscattering values $\langle I_i \rangle, \langle I_k \rangle$ on QY filter expression (Equation 1):

$$J_k(x,y) = \frac{\langle I_k \rangle}{N} \sum_{i=1}^{N} \frac{I_{i(x,y)}}{\langle I_i \rangle}, 1 \leq k \leq N \quad (1)$$

As this, four (4) different filters were tested for temporal and spatial filtering, being them: median 9x9, Frost 5x5, Frost 9x9 and Lee 3x3 (see [8] for a detailed description of every filter). All the 25 possibilities of filter (and no filter) combination were tested. Figure 3 shows the visual results of every tested combination over a deforested area:

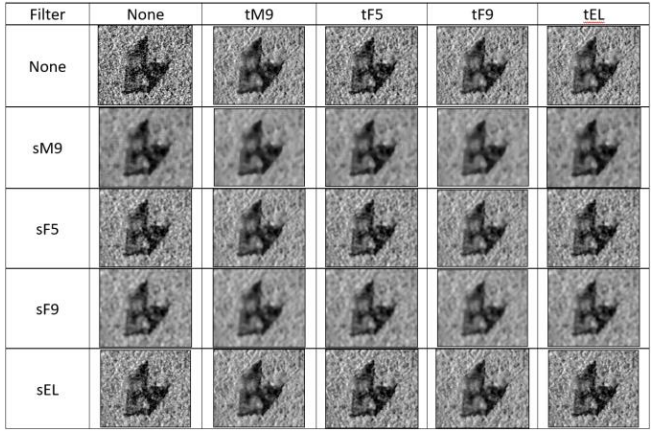

**Figure 3. Visual results of the filtering testing procedure over a deforested patch. The horizontal axis represents the different types of temporal filtering. The vertical axis represents the different types of spatial filters.**

Optimal filter selection was performed using the combination of two indexes: ENL (equivalent number of looks) and Range, a modified version of the target–to–clutter ratio (TCR) index [8]. The mathematical expressions of the used indexes are showed in Equation 2.

$$ENL = \frac{E[f]^2}{Var[f]}; \quad Range = 10 \log_{10} \frac{P_{90}^d}{P_{10}^d} \quad (2)$$

Being $f$ a despeckled image over a homogenous area and $P_{90}^d, P_{10}^d$ the 90-th and 10-th percentile of the time series distribution of the despeckled image over a deforested area.

Both indexes were calculated, for every filter combination, over 50 invariant forest locations (for *ENL*) and 50 deforested locations (for *Range*). Then the mean value of the indexes was normalized. The final score for every filter combination was taken as the mean of both indexes. Figure 4 shows the mean value of every combination score, for VV band. The optimal filter, as determined by this method, was a combination of a Median 9x9 filter inside a QY temporal filter and a Frost 9x9 spatial filter. VH showed similar results.

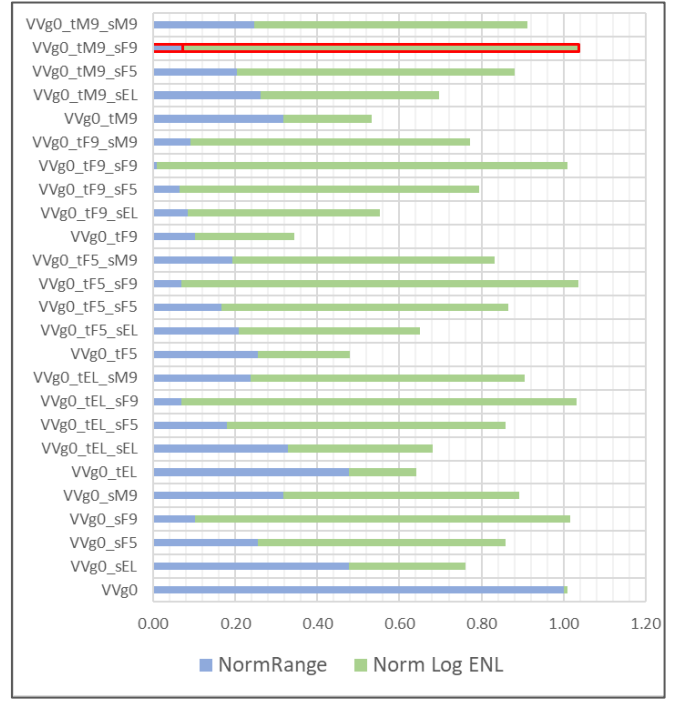

**Figure 4. Scores of the different tested filter combinations on the VV polarization. Median 9x9 temporal filter combined with a Frost 9x9 filter received the maximum score. VH results were quite similar.**

### 2.5 Filtered SAR data distribution over forest areas and threshold determination

To determine the statistical distribution of the filtered SAR data over intact forest, normality tests were performed over 50 locations sampled on the invariant forest data set. For every location, a 2 year-long log-scaled backscattering time series was extracted, and submitted to both Shapiro-Wilk (SW) and Kolmogorov-Smirnoff (KS) normality tests. The results show that in almost all the cases (86% SW, 100% KS, with $\alpha = 1\%$) it can be accepted that distributions are normal. Additionally, a Buttler test was performed to determine if all the sampled distributions had the same variance. The results were positive, with a 1% significance.

Assuming forest distribution of filtered SAR data to be normal with a known variance, stablishing a value for change detection thresholding becomes a simple unilateral statistical z-test, being the hypothesis:

$H_0: x = \mu$ ; $H_1: x < \mu$

The following Table 1 resumes the thresholding values obtained after resolving the z-test:

**Table 1. Thresholding values**

| Band | $Z_{crit}$ (α=5%) | $Z_{crit}$ (α=1%) |
|------|-------------------|-------------------|
| VV   | -0.8388           | -1.1864           |
| VH   | -1.2336           | -1.7447           |

## 3. RESULTS AND DISCUSSION

The application of the threshold values obtained for VV and VH distributions to generate deforestation alerts over samples of the deforested areas is illustrated on Figure 5:

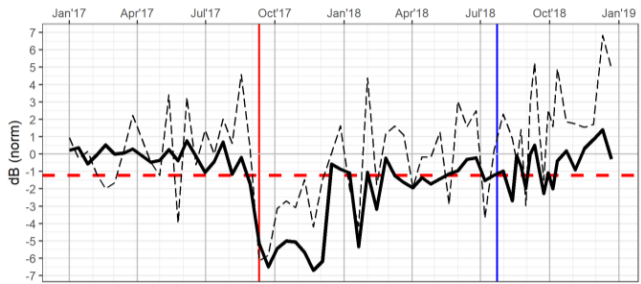

**Figure 5.** Filtered SAR response on a deforested time series, VH polarization (dotted line: non-filtered data, continuous line: filtered data). The red vertical line shows a confirmed alert (two sequential values under the threshold). The blue vertical line shows the DETER (optical) detected date.

The proposed automated alert system, triggered after having two sequential values under the threshold value, was systematically tested against 1000 and 100 locations on the deforested and intact forest data set, respectively. The results in terms of classification performance, for two different levels of significance, were as follows in Table 2.

**Table 2. Results of classification performance**

| Band | 5% significance | | | 1% significance | | |
|---|---|---|---|---|---|---|
| | CE | OE | MD | CE | OE | MD |
| VV | 64% | 5% | -84 days | 18% | 11% | -8 days |
| VH | 14% | 7% | -26 days | 0% | 17% | -4 days |

CE: Commission error; OE: Omission error; MD: Median delay between DETER and proposed alerts.

The results of the proposed alert system benchmarking, summarized on the table above, are very encouraging. The absence of false alerts when using VH polarization with a 1% significance is remarkable. The statistical characterization of the filtered SAR time series was very encouraging as well, as it shows that it is possible to model them using log-normal distributions.

## 4. CONCLUSIONS AND FUTURE WORK

Careful filter combination and statistical characterization were successfully used to determine statistically significant thresholds for change detection on SAR data.

The results point towards the feasibility of a basin-wide early warning system based on Sentinel-1 data. Future developments should try to explore the possibilities of multivariate gaussian statistics, to take advantage of the two available polarizations, and to reduce the detection errors.